# A Time Series Analysis-Based Forecasting Framework for the Indian Healthcare Sector

Jaydip Sen
Praxis Business School, Bakrahat Road, Off Diamond Harbor Road,
Kolkata – 700104, West Bengal, INDIA
email: jaydip.sen@acm.org

and

Tamal Datta Chaudhuri
Calcutta Business School, Diamond Harbour Road, Bishnupur – 743503
West Bengal, INDIA
email: tamalc@calcuttabusinessschool.org

**Abstract**

Designing efficient and robust algorithms for accurate prediction of stock market prices is one of the most exciting challenges in the field of time series analysis and forecasting. With the exponential rate of development and evolution of sophisticated algorithms and with the availability of fast computing platforms, it has now become possible to effectively and efficiently extract, store, process and analyze high volume of stock market data with diversity in its contents. Availability of complex algorithms which can execute very fast on parallel architecture over the cloud has made it possible to achieve higher accuracy in forecasting results while reducing the time required for computation. In this paper, we use the time series data of the healthcare sector of India for the period January 2010 till December 2016. We first demonstrate a decomposition approach of the time series and then illustrate how the decomposition results provide us with useful insights into the behavior and properties exhibited by the time series. Further, based on the structural analysis of the time series, we propose six different methods of forecasting for predicting the time series index of the healthcare sector. Extensive results are provided on the performance of the forecasting methods to demonstrate their effectiveness.

## 1. Introduction

Developing an accurate forecasting framework for predicting stock prices has been one of the most existing challenges to the researchers, particularly to those belonging to the Artificial Intelligence (AI) and Analytics community. Researchers working in this field have proposed various technical, fundamental and statistical indicators for predicting stock prices. These approaches have produced results with varying accuracy. In our previous work (Sen & Datta Chaudhuri, 2016a; Sen & Datta Chaudhuri, 2016b; Sen & Datta Chaudhuri, 2016c) we have proposed a novel approach towards portfolio diversification and prediction of stock prices. We argue that different sectors in an economy do not exhibit identical pattern of variations in their stock prices, and sectors differ from each other in terms of their trend pattern, their seasonal characteristics and also in the randomness in their time series. While the randomness aspect of stock market has been the major cornerstone of Efficient Market Hypothesis, the literature attempting to prove or disprove the hypothesis, has delved into various fundamental characteristics of different stocks and has come up with different results. We contend that besides the difference in the fundamental characteristics among stocks of different companies, performances of different stocks depend on the performance of the sectors to which they belong. Since each sector has its own factors responsible for its growth or sluggishness, the stocks belonging to different sectors are also influenced by these factors. The reasons behind the fortunes of the Information Technology (IT) sector in India are different from those of the metals sector of the healthcare sector, and these differences have to properly and quantitatively factored in for optimal portfolio choice and also for churning of the portfolio.

In this paper, we focus on the time series pattern of the healthcare sector in India. We take the time series index values of the Indian healthcare sector during the period January 2010 till December 2016. We decompose the time series using R programming language. We, then, demonstrate that the time series decomposition approach provides us with useful insights into various characteristics and properties of the time series. Using the trend, seasonal and random components values of the time series we can understand the growth pattern, the seasonal characteristics and the degree of randomness exhibited by the time series index values. We also propose an extensive framework for time series forecasting in which we present six different approaches of prediction of time series index values. We critically analyze the six approaches and also explain the reason why some methods perform better and produce lower values of forecast error in comparison to other methods.

The rest of the paper is organized as follows. Section 2 presents a brief literature survey and discusses some of the existing work on time series analysis and forecasting. Section 3 describes the methodology used in our proposed work for constructing various time series and decomposing the time series into its components. Section 4 depicts the results of decomposition of the healthcare sector time series index values into its trend, seasonal and random components. Based on the decomposition results, we explain several characteristics and behavior exhibited by the healthcare sector time series during the period of your study. Section 5 provides a detailed discussion on six different forecasting approaches that have been proposed in this work. Section 6 presents extensive results based on the application of each of the six forecasting techniques on the healthcare sector time series data. A comparative analysis of the techniques is also provided on the basis of five different metrics of the forecasting techniques – (i) maximum error, minimum error, mean error, standard deviation of error, and the *root mean square error* (RMSE). Finally, Section 7 concludes the paper.

## 2. Related Work

Several approaches and techniques have been proposed in the literature for forecasting of daily stock prices. Among these approaches, neural network-based approaches are extremely popular. Mostafa (2010) proposed a neural network-based technique for predicting stock market movements in Kuwait. Kimoto et al. (1990) presented a technique using neural network based on historical accounting data and various macroeconomic parameters to forecast variations in stock returns. Leigh et al. (2005) demonstrated methods of using linear regression and simple neural network models for predicting stock market indices in the New York Stock Exchange using stock exchange data for the period 1981-1999. Hammad et al. (2009) demonstrated how *artificial neural network* (ANN) models can be trained so that it converges and produces highly accurate results of forecasting of stock prices. Dutta et al. (2006) used ANN models for achieving highly accurate results of forecasting of Bombay Stock Exchange (BSE)'s SENSEX weekly closing values for the period of January 2002 - December 2003. Ying et al. (2009) used Bayesian Network (BN) – based approach to forecast stock prices of 28 companies listed in DJIA (Dow Jones Industrial Average) during 1988-1998. Tsai and Wang (2009) demonstrated that BN-based approaches usually produce higher accuracy in forecasting than traditional regression and neural network-based approaches. Tseng et al. (2012) presented a work in which the authors applied traditional *time series decomposition* (TSD), HoltWinters (H/W) models, Box-Jenkins (B/J) methodology and neural network-based approaches to 50 randomly selected stocks during the period September 1, 1998 till December 31, 2010 for forecasting future stock prices. The authors observed that forecasting errors are lower for B/J, H/W and normalized neural network model, while the errors are appreciably large for time series decomposition and non-normalized neural network model. Moshiri & Cameron (2010) designed a *back propagation network* (BPN) with econometric models to forecast inflation using (i) Box-Jenkins Autoregressive Integrated Moving Average (BJARIMA) model, (ii) Vector Autoregressive (VAR) model and (ii) Bayesian Vector Autoregressive (BVAR) model. Thenmozhi (2001) applied chaos theory for examining the pattern of changes of stock prices in BSE during the period August 1980 till September 1997. The author observed that the daily returns and the weekly returns of the BSE sensex exhibited nonlinearity characteristics and the time series of sensex movement was weakly chaotic. Hutchinson et al. (1994) proposed a novel approach using the principles of *learning networks* for estimating the price of a derivative.

ANN and Hybrid systems are particularly effective in forecasting stock prices for stock time series data. A large number of propositions based on ANN techniques for stock market prediction exists in the literature (Shen et al., 2007; Jaruszewicz & Mandziuk, 2004; Ning et al., 2009; Pan et al., 2005; Hamid & Iqbal, 2004; Chen et al., 2005; Chen et al., 2003; Hanias et al., 2007; de Faria et al., 2009). Many applications of hybrid systems in stock market time series data analysis have also been proposed in the literature (Wu et al., 2008; Wang & Nie, 2008; Perez-Rodriguez et al., 2005; Leung, et al., 2000; Kim, 2004).

In the literature, researchers have also proposed several forecasting techniques particularly focusing on various issues in the healthcare domain. Besseling & Shestalova (2011) propose a methodology to forecast medium-term expenditure in the healthcare sector in the Netherlands during the period 2011 – 2015. They identify several sub-sectors in the healthcare sector of the Netherlands and then decompose the healthcare expenditure into four categories: demographic, epidemiologic, budgetary and residuals. Based on this decomposition, they propose a novel forecasting approach for forecasting expenditure under the healthcare sector. Soyiri & Reidpath (2012) provide a broad framework for theoretical analysis and forecasting methods for health status. The authors discuss the key issues in health forecasting and properties of health data that influence the choice of a particular forecasting technique. Sense et al. (2015) propose a model that projects the evolution of supply of medical specials for three specific demand scenarios. The demand scenarios envisioned by the authors account for different drivers: demography, service utilization rates and hospital beds. Based on the model output the framework proposed by the authors uses a mixed integer programming model to determine the optimal assignment of medical specialization grants for different years under the period of study. Ostwald & Klingenberger (2016) proposed a novel approach for quantifying the economic significance of the oral healthcare sector in Germany. The authors presented a model for forecasting the growth in the oral healthcare sector based on various explanatory variables such as demographic change, take-up behavior, medical-technical progress, oral morbidity, aggregated supply, and income levels. Based on their study, the authors predict that by 2030, the healthcare sector in Germany will experience 19.2% increase in its gross value add. Finarelli & Johnson (2004) proposed a comprehensive, nine-step, quantitative demand forecasting model for healthcare services: (i) assemble historical data, (ii) analyze historical trends, (iii) identify key demand drivers, (iv) identify relevant benchmarks, (v) model existing conditions, (vi) develop core assumptions for population-based demand, (vii) develop core assumptions for provider-level demand, (viii) create a baseline forecast of future demand, (ix) test sensitivity of projections to changes in core assumptions. Cote & Tucker (2001) argue that four common methods of forecasting demand of health services are: (i) percent adjustment, (ii) 12-month moving average, (iii) trendline, and (iv) seasonalized forecast. Schweigler et al. (2009) investigated whether time series-based methods can make short-term forecasts of emergency department bed occupancy in hospitals. The authors observed that a sinusoidal model with *auto regression* (AR) – structured error term and a seasonal ARIMA model produce robust forecast in short-term time frame of 12 hours. Kadri et al. (2014) developed forecasting models for predicting daily attendances at the hospital emergency department in Lille, France during the period January 2012 till December 2012. The authors found that time series analysis based on ARIMA can be a very effective method for forecasting demands in hospital emergency services in short-term time frame of 12 hours. Beech (2001) proposed a method of deriving forecasts of the demand of market-based healthcare services from a broad range of available data. The data sets that the author used contained information on primary as well as secondary service areas, service-area populations by various demographic groupings, discharge utilization rates, market size and market share by service lines. The authors observed that market dynamics can allow development of trend models that can effectively forecast future demands. Myers and Green (2004) proposed a forecasting approach for predicting future demands in hospitals and developed a facility master plan based on the projected capacity.

In contrast to the work mentioned above, our approach in this paper is based on structural decomposition of time series of the healthcare sector index in India during the period January 2010 till December 2016. Based on the decomposition of the time series, we identified several interesting characteristics of the healthcare sector India. We particularly investigated the nature of the trend, seasonality pattern and degree of randomness of the time series. After analyzing the nature of the healthcare time series, we proposed six forecasting techniques for predicting the index values of the sector for each month of the year 2016. We have computed the relative accuracies of each of the forecasting techniques, and also have critically analyzed under what situations a particular technique performs better than the other techniques. Our proposed framework of analysis can be used as a broad approach for forecasting the behavior of other stock market indices in India.

## 3. Methodology

In this section, we provide a brief description about the methodology that we have followed in this work. We use *R programming language* (Ihaka & Gentleman, 1996) for all work of data management, data analysis and presentation of results. R is an open source language with a very rich set of libraries that makes it ideally suited for data analysis work. In this work, we use monthly data from the Bombay Stock Exchange (BSE) of India on the healthcare sector index for the period January 2010 till December 2016. The monthly index values of the healthcare sector for 7 years are stored in a plain text (.txt) file. The plain text file is then read into an R data object using the *scan( )* function R. The R data object is then converted into a time series object by using the *ts( )* function in R programming environment. The time series data object is then decomposed into three components – trend, seasonal and random – using the *decompose( )* function which is already defined in the TTR library in R. We plot the graphs of the healthcare time series data as well as its three components and make a detailed analysis of the behavior of the time series.

After a detailed analysis of the decomposition results of the time series of the healthcare sector, we propose six different approaches of forecasting time series values. For each of the propose forecasting method, we build the forecast model using the healthcare time series data for the period January 2010 till December 2015 and apply the model to forecast time series index values for each month of the year 2016. Since the actual values of the time series for all months of 2016 are known to us, we compute the error in forecasting using each method of forecast that we have proposed. A detailed comparative analysis, highlighting which method performs best under what situation and for what type of time series, is also presented.

In our previous work, we have highlighted the effectiveness of time series decomposition approach for robust analysis and forecasting of the Indian Auto sector (Sen & Datta Chaudhuri, 2016a; Sen & Datta Chaudhuri, 2016b). In another different work, we analyzed the behavior of two different sectors of Indian economy – the small cap sector and the capital goods sector – the former having a dominant random component while the latter exhibiting a significant seasonal component (Sen & Datta Chaudhuri, 2016c). Following a different approach of time series analysis, we also studied the behavior of the Indian Information Technology (IT) sector time series and the Indian Capital Goods sector time series (Sen & Datta Chaudhuri, 2016d). In another different work, we illustrated how time series decomposition-based approach enables us to check the consistency between the fund style and actual fund composition of a mutual fund (Sen & Datta Chaudhuri, 2016e).

In this work, we demonstrate how time series decomposition-based approach helps us in understanding the behavior and different properties of the healthcare time series of the Indian economy based on time series data for the period January 2010 till December 2016. We also investigate what forecasting approach is most effective for the healthcare time series. For this purpose, we compare several approaches of forecasting and identify the one that produces the minimum value of forecasting error. We critically analyze all the proposed forecasting approaches and explain why a particular approach has worked most effectively while some others have not done so for the healthcare time series data.

## 4. Time Series Decomposition Results

In this Section, we present the results that we have obtained in time series decomposition work of the healthcare index in BSE during the period January 2010 till December 2016. First, we create a plain text (.txt) file containing the monthly index values of the healthcare sector for the period January 2010 till December 2016. This file contained 84 records corresponding to the 84 months in the 7 years under study. We used the *scan( )* function in R language to read the text file and stored it in an R data object. Then, we converted this R data object into a time series object using the R function *ts( )*. We used the value of the *frequency* parameter in the *ts ( )* function as 12 so that the decomposition of the time series is carried out on monthly basis. After creating the time series data object, we used the function *plot( )* in R to draw the plot of the healthcare sector time series during the period January 2010 till December 2016. Figure 1 represents the graph of the healthcare sector time series.

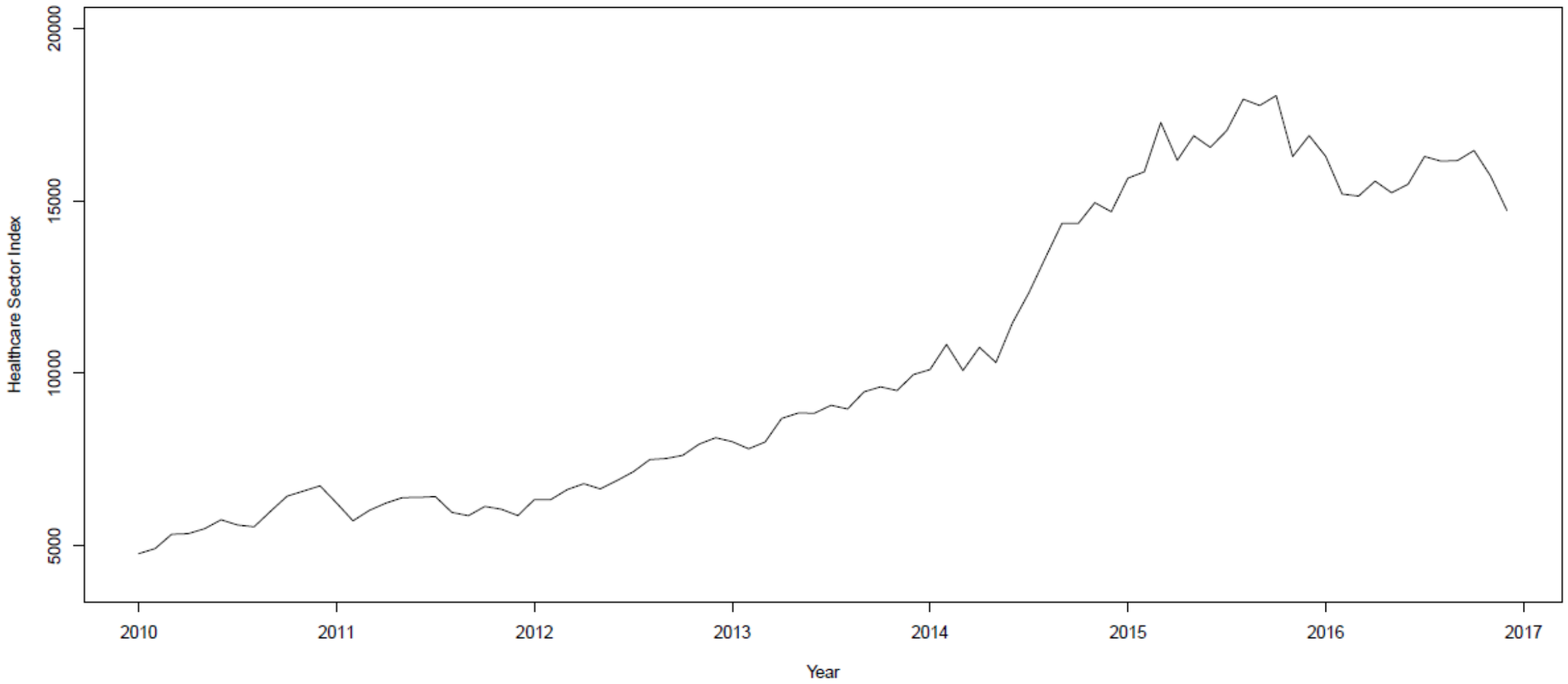

Figure1: Time series of healthcare sector index in India (Period: Jan 2010 – Dec 2016)

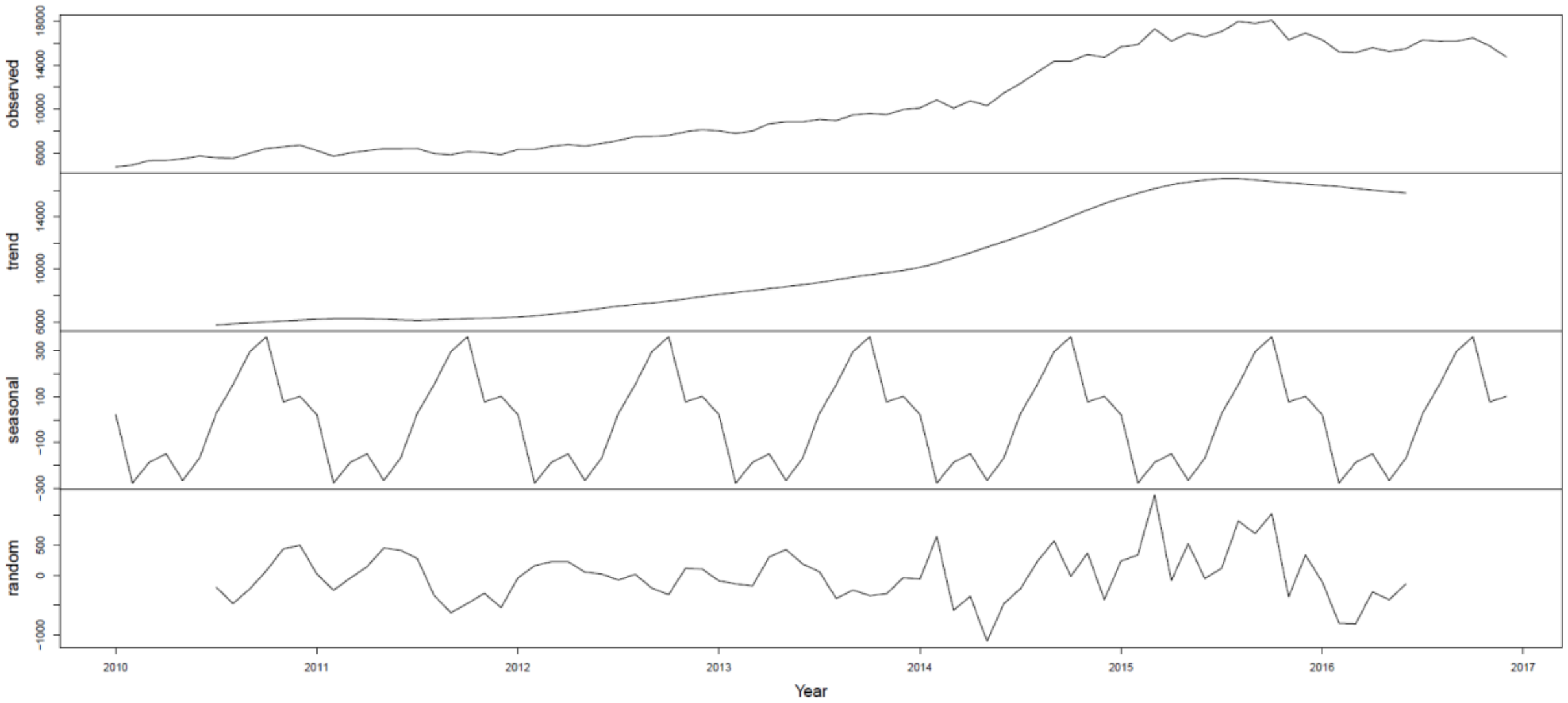

Figure 2: Decomposition of Healthcare sector index time series into trend, seasonal and random components (Period: Jan 2010 – Dec 2016)

In order to obtain a deeper insight into the behavior of the time series, we decomposed the time series object into its three components – trend, seasonal and random. We used the *decompose( )* function that is defined in the TTR library in R programming environment. The *decompose ( )* function is called with the healthcare time series object as its parameter and the three components of the time series are obtained. Figure 2 depicts the graphs of healthcare sector time series and its three components. Figure 2 consisted of four boxes arranged in a stack. The boxes display the overall time series, the trend, the seasonal and the random component respectively arranged from top to bottom in that order.

From Figure 1, it may be seen that the time series had an upward slope during the period 2010 till first couple of month in 2014. From mid 2014 till the end of the first quarter of 2015, the time series experienced an even steeper rise. However, it started falling after that and became sluggish in the entire year of 2016. Figure 2 shows the decomposition results of the healthcare time series. The three components of the time series are shown separately so that their relative behavior can be visualized.

Table 1 presents the numerical values of the time series data and its three components. The trend and the random components are not available for the period January 2010 – June 2010 and also for the period July 2016 – December 2016. This is due to the fact that trend computation requires long term data. Coughlan (2015) illustrated that the *decompose ( )* function in R uses a 12-month moving average method to compute the trend values in a time series. Hence, in order to compute trend values for the period January 2010 – June 2010, the *decompose ( )* function needs time series data for the period July 2009 – December 2009. Since, the data for the period July 2009 – December 2009 are not available in the dataset under our study, the trend values for the period January 2010 – June 2010 could not be computed. For similar reason, the trend values for the period July 2016 - December 2016, could not be computed due to non availability of the time series record for the period January 2017 – June 2017 in our dataset. It may be noted from Table 1 that the seasonal value for a given month remains constant throughout the entire period of study. For example, the seasonal component has a constant value of 22 for the month of January in every year from 2010 till 2016. It is interesting to note that due to the non-availability of the trend values for the periods January 2010 – June 2010 and July 2016 – December 2016, the random components for these periods could not also be computed by the *decompose( )* function. In other words, since the aggregate time series values are given by the sum of the corresponding trend, seasonal and random component values, and because of the fact that the seasonal value for a given month remains the same throughout, non-availability of trend values for a period makes the random components values also unavailable for the same period.

Table 1: Healthcare sector index time series and its components (Jan 2010 – Dec 2016)

| Year | Month | Aggregate Index | Trend | Seasonal | Random |
|---|---|---|---|---|---|
| **2010** | January | 4765 | | 22 | |
| | February | 4913 | | -277 | |
| | March | 5328 | | -186 | |
| | April | 5345 | | -148 | |
| | May | 5490 | | -265 | |
| | June | 5749 | | -167 | |
| | July | 5597 | 5768 | 28 | -198 |
| | August | 5544 | 5863 | 153 | -472 |
| | September | 5996 | 5925 | 297 | -226 |
| | October | 6433 | 5991 | 363 | 79 |
| | November | 6583 | 6066 | 78 | 440 |
| | December | 6734 | 6130 | 102 | 501 |
| | January | 6237 | 6192 | 22 | 23 |
| | February | 5718 | 6244 | -277 | -249 |
| **2011** | March | 6024 | 6256 | -186 | -45 |
| | April | 6233 | 6238 | -148 | 143 |
| | May | 6393 | 6204 | -265 | 454 |
| | June | 6398 | 6146 | -167 | 420 |
| | July | 6421 | 6114 | 28 | 280 |
| | August | 5962 | 6144 | 153 | -335 |
| | September | 5868 | 6195 | 297 | -624 |
| | October | 6136 | 6243 | 363 | -470 |
| | November | 6055 | 6277 | 78 | -300 |
| | December | 5871 | 6308 | 102 | -539 |
| **2012** | January | 6336 | 6358 | 22 | -44 |
| | February | 6336 | 6452 | -277 | 161 |
| | March | 6626 | 6585 | -186 | 227 |
| | April | 6796 | 6716 | -148 | 228 |
| | May | 6645 | 6857 | -265 | 53 |
| | June | 6884 | 7030 | -167 | 22 |
| | July | 7142 | 7194 | 28 | -80 |
| | August | 7496 | 7325 | 153 | 17 |
| | September | 7528 | 7444 | 297 | -213 |
| | October | 7620 | 7581 | 363 | -324 |
| | November | 7946 | 7752 | 78 | 117 |
| | December | 8132 | 7925 | 102 | 104 |
| **2013** | January | 8017 | 8087 | 22 | -93 |
| | February | 7810 | 8229 | -277 | -142 |
| | March | 8008 | 8371 | -186 | -177 |
| | April | 8691 | 8535 | -148 | 305 |
| | May | 8847 | 8682 | -265 | 430 |
| | June | 8845 | 8823 | -167 | 189 |
| | July | 9074 | 8987 | 28 | 59 |
| | August | 8966 | 9200 | 153 | -388 |
| | September | 9464 | 9413 | 297 | -246 |
| | October | 9609 | 9586 | 363 | -340 |
| | November | 9501 | 9733 | 78 | -310 |
| | December | 9966 | 9903 | 102 | -40 |
| **2014** | January | 10110 | 10148 | 22 | -61 |
| | February | 10840 | 10468 | -277 | 649 |
| | March | 10084 | 10854 | -186 | -584 |
| | April | 10757 | 11256 | -148 | -350 |
| | May | 10315 | 11681 | -265 | -1101 |
| | June | 11462 | 12105 | -167 | -476 |
| | July | 12341 | 12533 | 28 | -220 |
| | August | 13357 | 12974 | 153 | 230 |
| | September | 14352 | 13483 | 297 | 572 |
| | October | 14354 | 14009 | 363 | -18 |
| | November | 14957 | 14510 | 78 | 370 |
| | December | 14693 | 14997 | 102 | -406 |
| **2015** | January | 15667 | 15405 | 22 | 239 |
| | February | 15855 | 15793 | -277 | 338 |
| | March | 17285 | 16128 | -186 | 1343 |
| | April | 16187 | 16426 | -148 | -90 |
| | May | 16900 | 16636 | -265 | 529 |
| | June | 16564 | 16784 | -167 | -53 |
| | July | 17048 | 16903 | 28 | 117 |
| | August | 17962 | 16903 | 153 | 906 |
| | September | 17779 | 16787 | 297 | 695 |
| | October | 18066 | 16672 | 363 | 1031 |
| | November | 16298 | 16578 | 78 | -358 |
| | December | 16905 | 16465 | 102 | 338 |
| **2015** | January | 16305 | 16389 | 22 | -106 |
| | February | 15208 | 16283 | -277 | -798 |
| | March | 15149 | 16141 | -186 | -806 |
| | April | 15582 | 16008 | -148 | -278 |
| | May | 15246 | 15918 | -265 | -407 |
| | June | 15493 | 15804 | -167 | -144 |
| | July | 16299 | | 28 | |
| | August | 16162 | | 153 | |
| | September | 16181 | | 297 | |
| | October | 16472 | | 363 | |
| | November | 15734 | | 78 | |
| | December | 14728 | | 102 | |

From Table 1, some important observations may be made. First, we can see that seasonal component has the maximum value of 363 in the month of October, while the lowest value of seasonality -265 is observed in the month of February. The seasonal component is found to have high positive values during the months of August till December, while negative values of seasonality are observed during the months of February till June. In order to analyze the impact of seasonality on the healthcare time series, we computed some statistics on the seasonal component values. We computed the percentage contribution of the seasonal component on the aggregate time series values and found the following. The maximum, the minimum, the mean of the absolute values, and the standard deviation of the percentage of the seasonal components with respect to the aggregate time series values were found to be 5.92, -4.84, 1.95 and 2.44 respectively. The maximum percentage of seasonal component was found in the month of October 2011, while the minimum was observed in the month of February 2011. While the low value of the mean percentage indicated that the time series was not seasonal, the high value of standard deviation in comparison to the mean value implied that the seasonal percentages exhibited high level of dispersion among themselves.

Second, the trend component of the time series was very sluggish during the period July 2010 till December 2011. However, the trend started increasing at a faster rate from January 2012 till May 2014. The rate of increase in trend value became even faster from May 2014 and it continued till July 2015. However, from July 2015 till the end of our period of study (i.e., till June 2016), the trend became sluggish and started falling in its magnitude at a very slow rate. The maximum, the minimum, the mean, and the standard deviation of the percentage of the trend component with respect to the aggregate time series were found to be 113.24, 91.03, 100.35 and 4.23 respectively indicating that the trend was the single most dominant component in the time series. The maximum percentage of trend component was found in the month of May 2014, while the minimum was found in the month of December 2010.

Third, the maximum and the minimum values of the random component of the time series were found to be 1343 and -1101 respectively. These values are quite modest in comparison to the aggregate time series values. In order to understand the contribution of the random component on the overall time series, we computed the maximum, the minimum, the mean of the absolute values and the standard deviation of the percentage of random component values with respect to the aggregate time series values. These values were found to be 7.77, -10.67, 3.37 and 4.23 respectively. It indicated that while the random component is not dominant in the time series, the values of the random component exhibited large deviations across their mean value. The random component contributed its maximum percentage to the aggregate time series in the month of March 2015, while the lowest percentage was found in the month of May 2014.

The overall conclusion is that the healthcare time series is primarily dominated by its trend component, while seasonal and random components are having not significant contributions to the aggregate time series. However, the seasonal and random components exhibited significant variations across their mean values.

## 5. The Proposed Forecasting Methods

In this Section, we present some forecasting methods that we have applied on the time series data of the healthcare sector index. We propose six different approaches to forecasting and also present the performance of these approaches on the healthcare sector time series data. For the purpose of comparative analysis of different approaches of forecasting, we use five different metrics and identify which method leads to lowest value of forecasting error. We also critically analyze the approaches and argue why one method perform better than the others on the given dataset of healthcare sector time series index for the period January 2010 – December 2016. In this Section, we describe the six approaches and in Section 6, we provide the detailed forecasting results as these approaches are applied on the healthcare sector dataset.

**Method 1:** In this method, we use the healthcare sector time series data for the period January 2010 till December 2015 for the purpose of forecasting the monthly index values for each month of the year 2016. We use *HoltWinters( )* function in R library *forecast* for this purpose. In order to make a robust forecasting, we use *HoltWinters* model with a changing trend and an additive seasonal component that best fits the healthcare time series index data. The forecast *horizon* in the *HoltWinters* model has been chosen to be 12 so that the forecasted values for all months of 2016 can be obtained by using the method at the end of the year 2015. Forecast error is computed for each month of 2016 and an overall RMSE value is also derived for this method.

**Method II:** In this approach, forecasting of the time series index for the healthcare sector for each month of the year 2016 is done using a forecast horizon of 1 month. For example, for the purpose of forecasting the index for the month of March 2016, time series data from January 2010 till February 2016 are used to develop the forecasting model. We use the *HoltWinters* model with a changing trend and an additive seasonal component with a forecast horizon of 1 month. Since the forecast horizon is small, the model is likely to produce higher accuracy in forecasting compared to the approach followed in Method I that used a forecast horizon of 12 months. The forecast error corresponding to each month of 2016 and an overall RMSE value for the model is computed.

**Method III:** In this method, we first use the time series data for the healthcare sector index for the period January 2010 till December 2015 and derive the trend and the seasonal component values. This method yields the values of the trend component of the time series for the period from July 2010 till June 2015. Based on these computed trend values for the period January 2010 till June 2015, we make forecast for the trend values for the period July 2015 till June 2016, using the *HoltWineters( )* function in R with a changing trend component level but without any seasonality component. In other words, in the *HoltWinters( )* function in R, we set the parameter '*beta*' = TRUE and the parameter '*gamma*' = FALSE, for the purpose of forecasting the trend values. The forecasted trend values are added to the seasonal component values of the corresponding months (based on the time series data for the period January 2010 till December 2015) to arrive at the forecasted aggregate of the trend and seasonal components. Now, we consider the time series of the healthcare sector index values for the entire period, i.e., from January 2010 till December 2016, and decompose it into its trend, seasonal and random components. Based on this time series, we compute the aggregate of the actual trend and the actual seasonal component values for the period July 2015 till June 2016. We derive the forecasting accuracy of this approach by calculating the percentage of deviation of the aggregate of the actual trend and the actual seasonal component values with respect to the corresponding aggregate values of the forecasted trend and past seasonal components for each month during July 2015 to June 2016. An overall RMSE value is also computed.

**Method IV:** The approach followed in this method is exactly similar to that used in Method III. However, unlike Method III that used *HoltWinters( )* function with changing trend component and a nil seasonal component, this method uses a linear regression model for the purpose of forecasting the trend component values for the period from July 2015 till June 2016. The *lm ( )* function in R is used for building the linear regression model with trend component as the response variable and time as the predictor variable. The regression model is built using the trend values for the period July 2010 till June 2015. The aggregate of the predicted trend values and the past seasonal values are compared with the aggregate of the actual trend values and the actual seasonal values for the period from July 2015 till June 106. The error in forecasting and an overall RMSE value is computed as in Method III.

**Method V:** We use *Auto Regressive Integrated Moving Average* (ARIMA) based approach of forecasting in this method. For the purpose of building the ARIMA model, we use the healthcare sector time series data for the period January 2010 till December 2015. Based on this training data set and using the *auto.arima( )* function defined in the forecast package in R, we determine the three parameters of the *Auto Regressive Moving Average* (ARMA) model, i.e. the *Auto Regression* parameter (*p*), the *Difference* parameter (*d*), and the *Moving Average* parameter (*q*). We construct the ARIMA model of forecasting using the *arima( )* function in R with the two parameters as: (i) healthcare time series R object (based on data for the period from January 2010 till December 2015), (ii) the order of the ARMA i.e., (*p*, *d*, *q*).Using the resultant ARIMA model, we call the function *forecast.Arima( )* with parameters: (i) ARIMA model object, and (ii) forecast horizon = 12 months (in this approach). Since we use a forecast horizon of 12 months, we can compute the forecasted values of each month of the year 2016 at the end of the year 2015. The error in forecasting for each month of 2016 and the RMSE value of this method are also computed.

**Method VI:** Similar to Method V, this approach also is based on an ARIMA model. However, unlike Method V that used forecast horizon of 12 months, this method uses a forecast horizon of 1 month. For the purpose of forecasting, the ARIMA model is built using time series data for the period January 2010 till the month previous to the month for which forecasting is being made. For example, for the purpose of prediction of the time series index for the month of May 2016, the time series data form January 2010 till April 2016 is used for building the ARIMA model. Since the training data set for building the ARIMA model constantly changes in this approach, we evaluate the ARIMA parameters (i.e., *p*, *d*, and *q*) before every round of forecasting. In other words, for each month of the year 2016, before we make the forecast for the next month, we compute the values of the parameters of the ARIMA model.

## 6. Forecasting Results

In this Section, we provide results on the performance of the six forecasting methods that we have described in Section 5. For each method, we computed the RMSE value so that the methods can be compared on the basis of their forecasting accuracies.

**Method I:** The results obtained using these methods are presented in Table 2. In Figure 3, we have plotted the actual values of the healthcare sector index and the corresponding predicted values for each month of the year 2016.

Table 2: Computation results using Method I of forecasting

| Month (A) | Actual Index (B) | Forecasted Index (C) | % Error (C-B)/B *100 | RMSE |
|---|---|---|---|---|
| Jan | 16305 | 17725 | 8.71 | 2863 |
| Feb | 15208 | 17799 | 17.04 | |
| Mar | 15149 | 18106 | 19.52 | |
| Apr | 15582 | 17544 | 12.59 | |
| May | 15246 | 17994 | 18.02 | |
| Jun | 15493 | 18160 | 17.21 | |
| Jul | 16299 | 18784 | 15.25 | |
| Aug | 16162 | 19331 | 19.61 | |
| Sep | 16181 | 19215 | 18.75 | |
| Oct | 16472 | 19238 | 16.79 | |
| Nov | 15734 | 18560 | 17.96 | |
| Dec | 14728 | 19341 | 31.32 | |

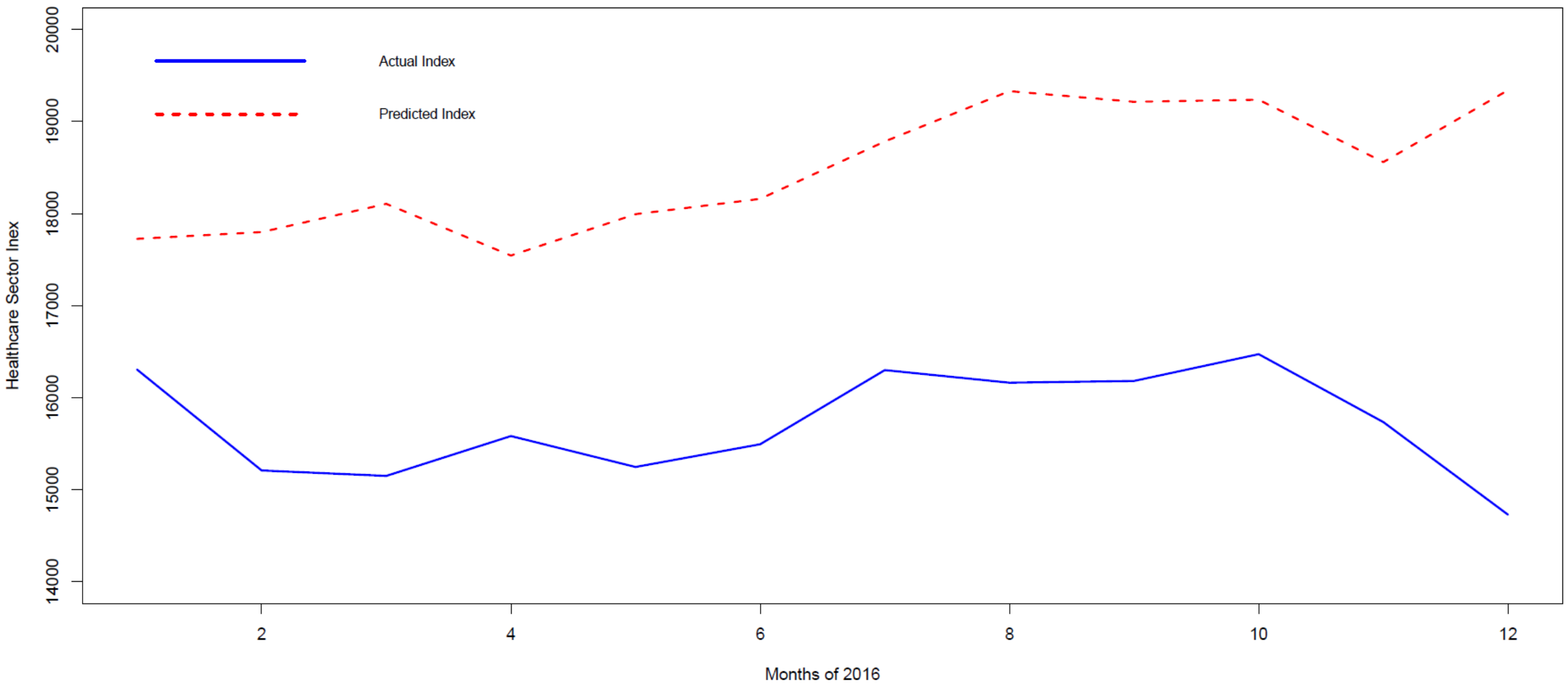

Figure 3. Actual and predicted values of healthcare sector index using Method I of forecasting (Period: Jan 2016 – Dec 2016)

**Observations:** We observe from Table 2 that the forecasted values deviate substantially from the actual values for most of the months in the year 2016. For each of the 12 months, this method has yielded forecasted value that is higher than the corresponding actual index value. One reason for such high error values in this method may be attributed to the long forecast horizon (12 months) used in this method for this method. We also see that the highest value of error percentage is observed in the month of December 2016. While the actual index decreased in its value during the period October – December 2016, the forecasted index increased in its value for the period November – December 2016. Hence, for the month of December 2016, a high percentage value of error (31.32%) is produced.

**Method II:** The results of forecasting using Method II are presented in Table 3. In Figure 4, the actual index values and their corresponding predicted values are plotted.

Table 3: Computation results using Method II of forecasting

| **Month** | **Actual Index** | **Forecasted Index** | **% Error** | **RMSE** |
|---|---|---|---|---|
| **(A)** | **(B)** | **(C)** | **(C-B)/B *100** | 894 |
| Jan | 16305 | 17725 | 8.71 | |
| Feb | 15208 | 16709 | 9.87 | |
| Mar | 15149 | 15622 | 3.12 | |
| Apr | 15582 | 15481 | 0.65 | |
| May | 15246 | 15884 | 4.18 | |
| Jun | 15493 | 15398 | 0.61 | |
| Jul | 16299 | 15161 | 6.98 | |
| Aug | 16162 | 16151 | 0.07 | |
| Sep | 16181 | 16515 | 2.06 | |
| Oct | 16472 | 16623 | 0.92 | |
| Nov | 15734 | 16622 | 5.64 | |
| Dec | 14728 | 16297 | 10.65 | |

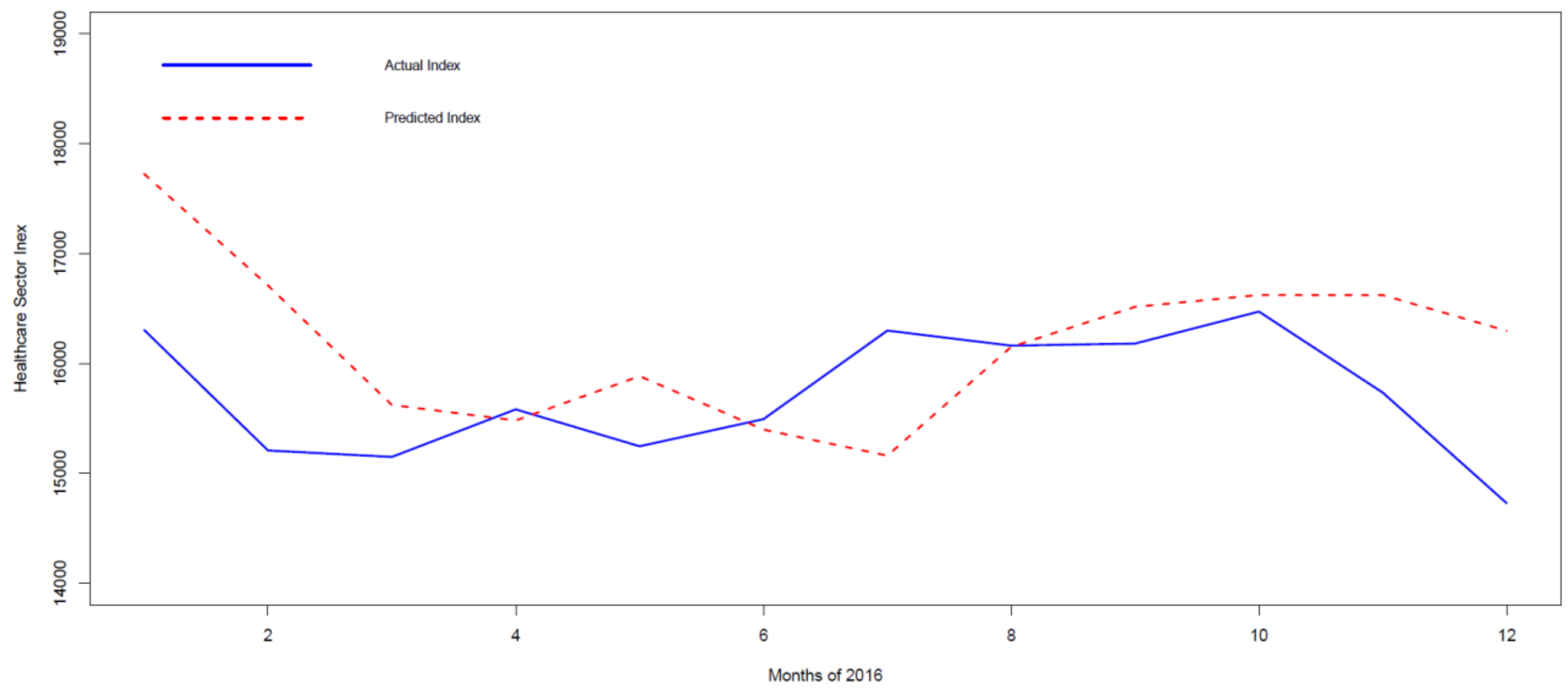

Figure 4. Actual and predicted values of healthcare sector index using Method II of forecasting (Period: Jan 2016 – Dec 2016)

**Observations:** We observe from Table 3 and Figure 4 that the forecasted values very closely match with the actual values of the time series index. The lowest value of error percentage is found to be 0.07 which occurred in the month of August 2016 and the highest error percentage value of 10.65 was found in the month of December 2016. The RMSE value for this method is 894, which can be considered as quite low. This clearly demonstrates that *HolWinters* additive model with a prediction horizon of 1 month can very effectively and accurately forecast future time series values.

Table 4: Computation results using Method III of forecasting

| **Month** | **Actual Trend** | **Actual Seasonal** | **Actual (Trend + Seasonal)** | **Forecasted Trend** | **Past Seasonal** | **Forecasted (Trend + Seasonal)** | **% Error** | **RMSE** |
|---|---|---|---|---|---|---|---|---|
| **A** | **B** | **C** | **D** | **E** | **F** | **G** | **(G-D)/D *100** | |
| Jul | 16903 | 28 | 16931 | 16932 | 7 | 16939 | 0.05 | 1782 |
| Aug | 16903 | 153 | 17056 | 17080 | 25 | 17105 | 0.29 | |
| Sep | 16787 | 297 | 17084 | 17228 | 161 | 17389 | 1.79 | |
| Oct | 16672 | 363 | 17035 | 17376 | 160 | 17536 | 2.94 | |
| Nov | 16578 | 78 | 16656 | 17524 | 152 | 17676 | 6.12 | |
| Dec | 16465 | 102 | 16567 | 17672 | 38 | 17710 | 6.90 | |
| Jan | 16389 | 22 | 16411 | 17820 | 47 | 17867 | 8.87 | |
| Feb | 16283 | -277 | 16006 | 17968 | 114 | 18082 | 12.97 | |
| Mar | 16141 | -186 | 15955 | 18116 | 22 | 18138 | 13.68 | |
| Apr | 16008 | -148 | 15860 | 18264 | 90 | 18354 | 15.73 | |
| May | 15918 | -265 | 15653 | 18412 | 180 | 18592 | 18.78 | |
| Jun | 15804 | -167 | 15637 | 18560 | 135 | 18695 | 19.56 | |

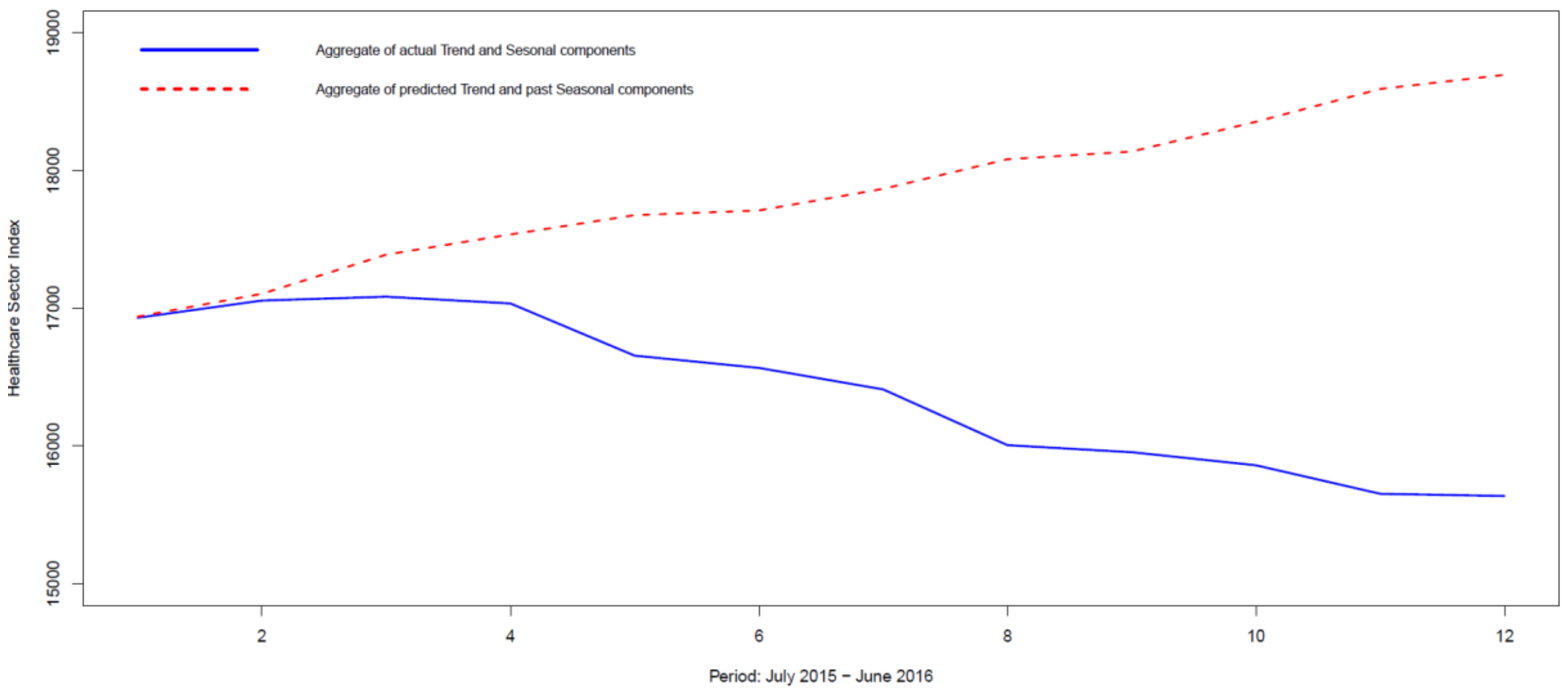

Figure 5. Actual and predicted values of the sum of trend and seasonal components of healthcare sector index using Method III of forecasting (Period: Jan 2016 – Dec 2016)

**Method III:** The results of forecasting using this method are presented in Table 4. Figure 5 depicts the actual index values and their corresponding predicted values for all months of the year 2016 using Method III of forecasting. The actual trend and seasonal component values for the period from July 2015 till June 2016 (computed based on the time series data for the period from January 2010 till December 2016) and their aggregated monthly values are noted in Columns *B*, *C* and *D* respectively in Table 4. The forecasted trend values (using *HoltWinters* method with changing trend component and nil seasonal component and with a forecast horizon of 12 months) and the past seasonal component values (based on time series data for the period from January 2010 till December 2015) and their corresponding aggregate values are noted in columns *E*, *F* and *G* respectively in Table 4. The error value for each month and an overall RMSE value for this method are also computed.

**Observation:** It may be observed from Table 4 and Figure 5 that the error in forecasting consistently increased from a very small value of 0.05 percent in July 2016 to a moderate value of 19.56 percent in June 2016. The computed RMSE value of 1782 indicated that overall error in forecasting is quite moderate with respect to the mean value of the actual index which is 16404. A causal look at Figure 5 indicates that while the actual aggregate of the trend and the seasonal component values consistently decreased over the entire period from July 2015 till June 2016, the aggregate of the predicted trend and the past seasonal component has a positive gradient throughout the entire period. The trend component is predicted using *HoltWinters* method with a changing trend parameter thereby producing predicted trend that always has a positive gradient. The increasing pattern of the predicted trend on being added with the seasonal component produces aggregate predicted values of trend and seasonal component that have a consistent upward slope. Since the aggregate of the actual values of trend and seasonal component has consistently decreased, the difference between the predicted values of the aggregate of trend and seasonal components and their corresponding actual values consistently increased with time. This resulted into the highest value of error in forecasting for the month of December 2016.

**Method IV:** Table 5 presents the results of forecasting using Method IV, while Figure 6 shows the actual index values and their corresponding predicted values for all months during the period July 2015 till June 2016. The actual trend and seasonal component values for the period from July 2015 till June 2016 (computed based on the time series data for the period from January 2010 till December 2016) and their aggregated monthly values are noted in Columns *B*, *C* and *D* respectively in Table 4. The forecasted trend values (using a linear regression method between index value and time for 12 months) and the past seasonal component values (based on time series data for the period from January 2010 till December 2015) and their corresponding aggregate values are noted in columns *E*, *F* and *G* respectively in Table 5. The error value for each month and an overall RMSE value for this method are also computed.

Table 5: Computation results using Method IV of forecasting

| Month | Actual Trend | Actual Seasonal | Actual (Trend + Seasonal) | Forecasted Trend | Past Seasonal | Forecasted (Trend + Seasonal) | % Error | RMSE |
|---|---|---|---|---|---|---|---|---|
| **A** | **B** | **C** | **D** | **E** | **F** | **G** | **(G-D)/D *100** | |
| Jul | 16903 | 28 | 16931 | 14660 | 7 | 14667 | 13.37 | 1339 |
| Aug | 16903 | 153 | 17056 | 14841 | 25 | 14866 | 12.84 | |
| Sep | 16787 | 297 | 17084 | 15021 | 161 | 15182 | 11.13 | |
| Oct | 16672 | 363 | 17035 | 15202 | 160 | 15362 | 9.82 | |
| Nov | 16578 | 78 | 16656 | 15382 | 152 | 15534 | 6.74 | |
| Dec | 16465 | 102 | 16567 | 15563 | 38 | 15601 | 5.83 | |
| Jan | 16389 | 22 | 16411 | 15743 | 47 | 15790 | 3.78 | |
| Feb | 16283 | -277 | 16006 | 15924 | 114 | 16038 | 0.20 | |
| Mar | 16141 | -186 | 15955 | 16104 | 22 | 16126 | 1.07 | |
| Apr | 16008 | -148 | 15860 | 16285 | 90 | 16375 | 3.25 | |
| May | 15918 | -265 | 15653 | 16465 | 180 | 16645 | 6.34 | |
| Jun | 15804 | -167 | 15637 | 16646 | 135 | 16781 | 7.32 | |

**Observation:** It is clear from Table 4 and Figure 5 that the pattern of error in Method IV is different from that of Method III. The error value in July 2015 was 13.5 percent and it consistently decreased till

February 2016 reaching a negligible value of 0.20 per cent. The error started increasing at a very low rate since March 2016 reaching a value of 7.32 per cent in June 2016. The computed RMSE value for this method is 1339 which is only 8.16 per cent of the mean value of the actual index which is 16404 indicating that the method of forecasting is quite effective. A careful look at Figure 6 enables one to see that the forecasted aggregate of trend and seasonal component using Method IV consistently increased over the entire period from July 2015 till June 2016 as it was also the case in Method III. However, unlike Method III, the difference between the actual and the forecasted values of the aggregate of the trend and the seasonal components decreased consistently for the first eight months and then increased very moderately over the last four months under the period of study. This made Method IV more accurate than Method III in its forecasting results which is also manifested in its lower value of RMSE.

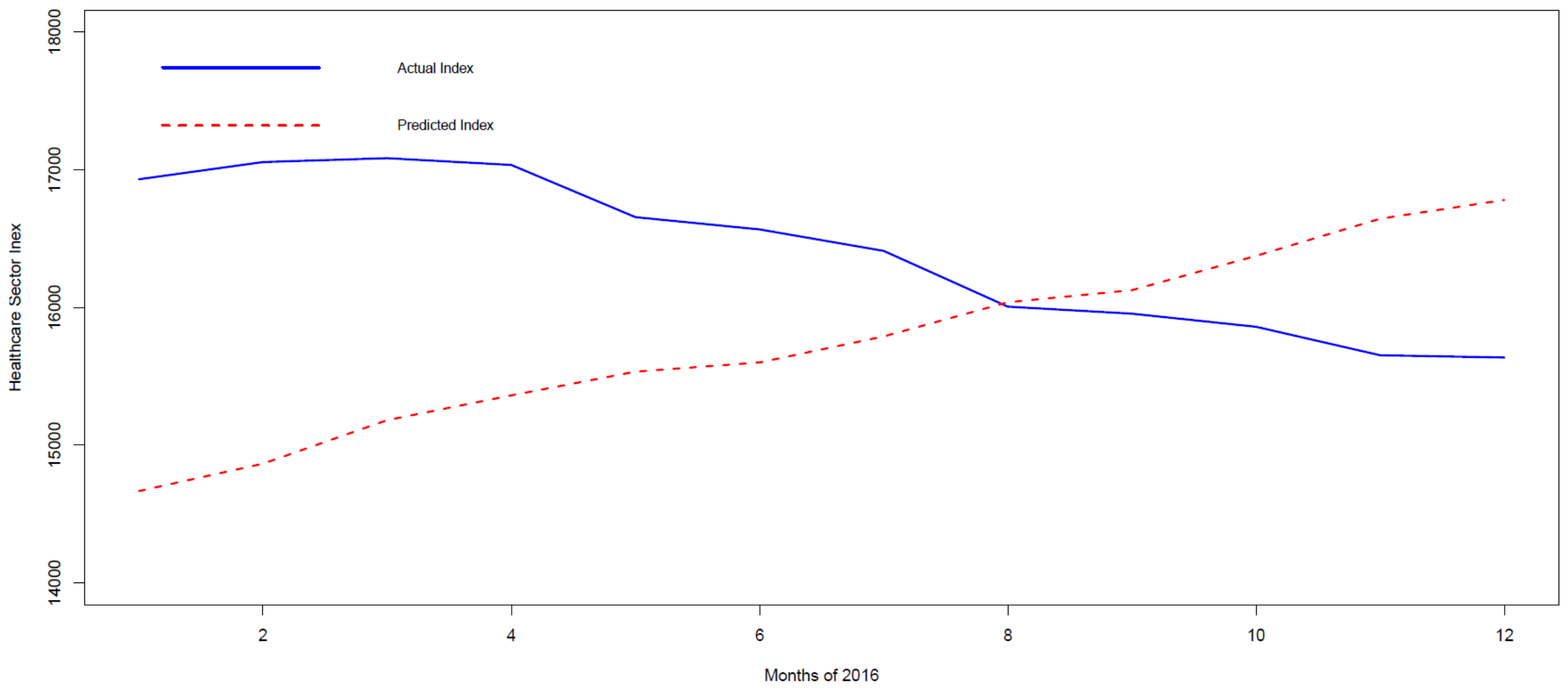

Figure 6. Actual and predicted values of the sum of trend and seasonal components of healthcare sector index using Method IV of forecasting (Period: Jan 2016 – Dec 2016)

**Method V:** Applying *auto.arima( )* function on the healthcare sector time series index values for the period from January 2010 till December 2015, we obtain the parameter values for the time series as: $p$ = 1, $d$ = 1, and $q$ = 2. Therefore, the auto sector time series for the period from January 2010 till December 2014 is designed as an ARMA (1, 1, 2) model. We also compute the *partial auto correlation function* (PACF) and the *auto correlation function* (ACF) and plot these functions so as to cross-check the values of the parameters $p$ and $q$. From Figure 7, it is evident that the minimum integral lag beyond which all partial autocorrelation values are insignificant is 1. Hence, it is confirmed that the value of the parameter $p$ is 1. Figure 8 shows that minimum integral value of lag beyond which all autocorrelation values are insignificant is 2. Hence the parameter $q$ has a value of 2. We also cross-checked the value of the parameter $d$ = 1, by plotting the first order difference of the time series and checking that the first order difference is stationary. Similarly Using the *arima( )* function with its two parameters: (i) the healthcare sector time series R object and (ii) the order (1, 1, 2) of ARMA, we construct the ARIMA model. Finally, we use the function *forecast.Arima( )* with two parameters: (i) the ARIMA model and (ii) the time horizon of forecast = 12 months, for forecasting the index values of the time series for all the twelve months of the year 2016. Table 6 presents the results of forecasting using this method, while Figure 9 depicts the actual values and predicted index values for all months in the year 2016.

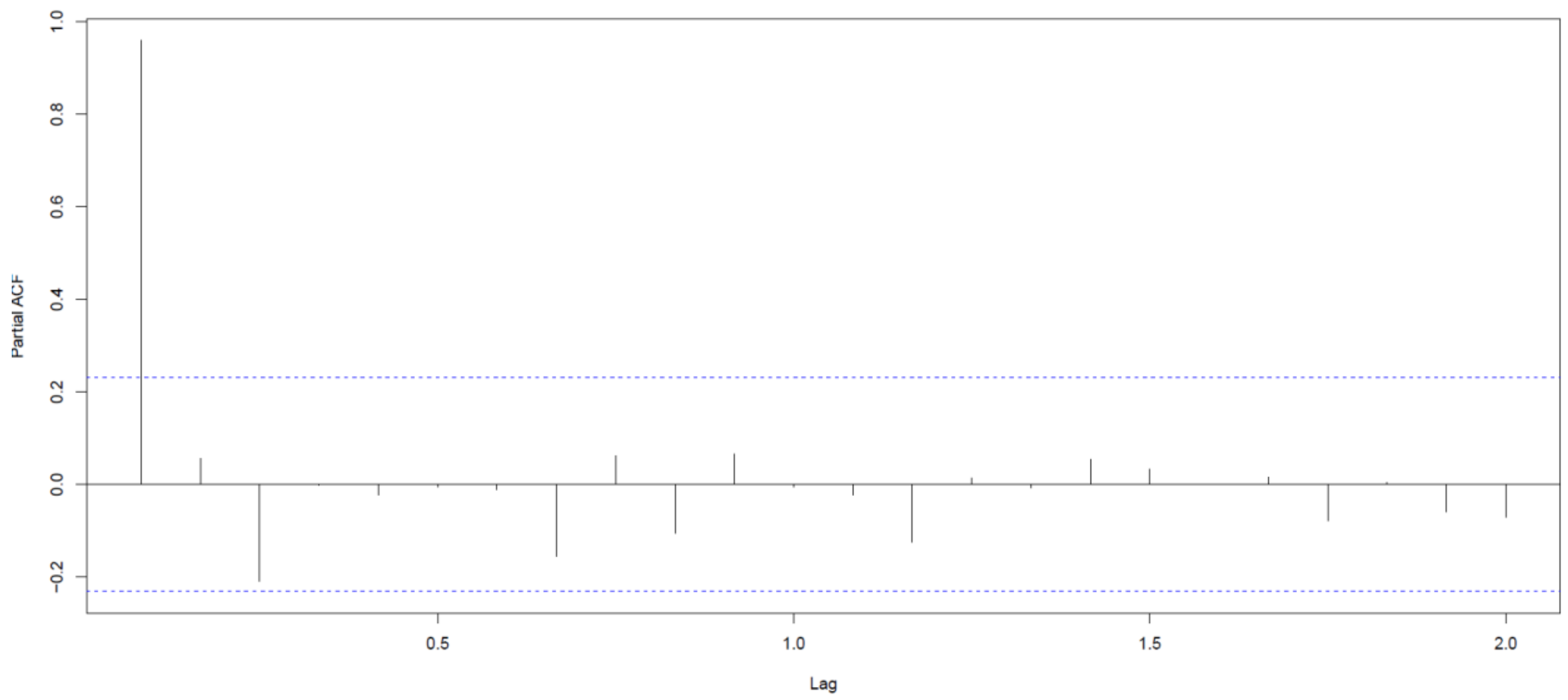

Figure 7. Partial auto correlation function of the healthcare time series (Period: Jan 2010 – Dec 2015)

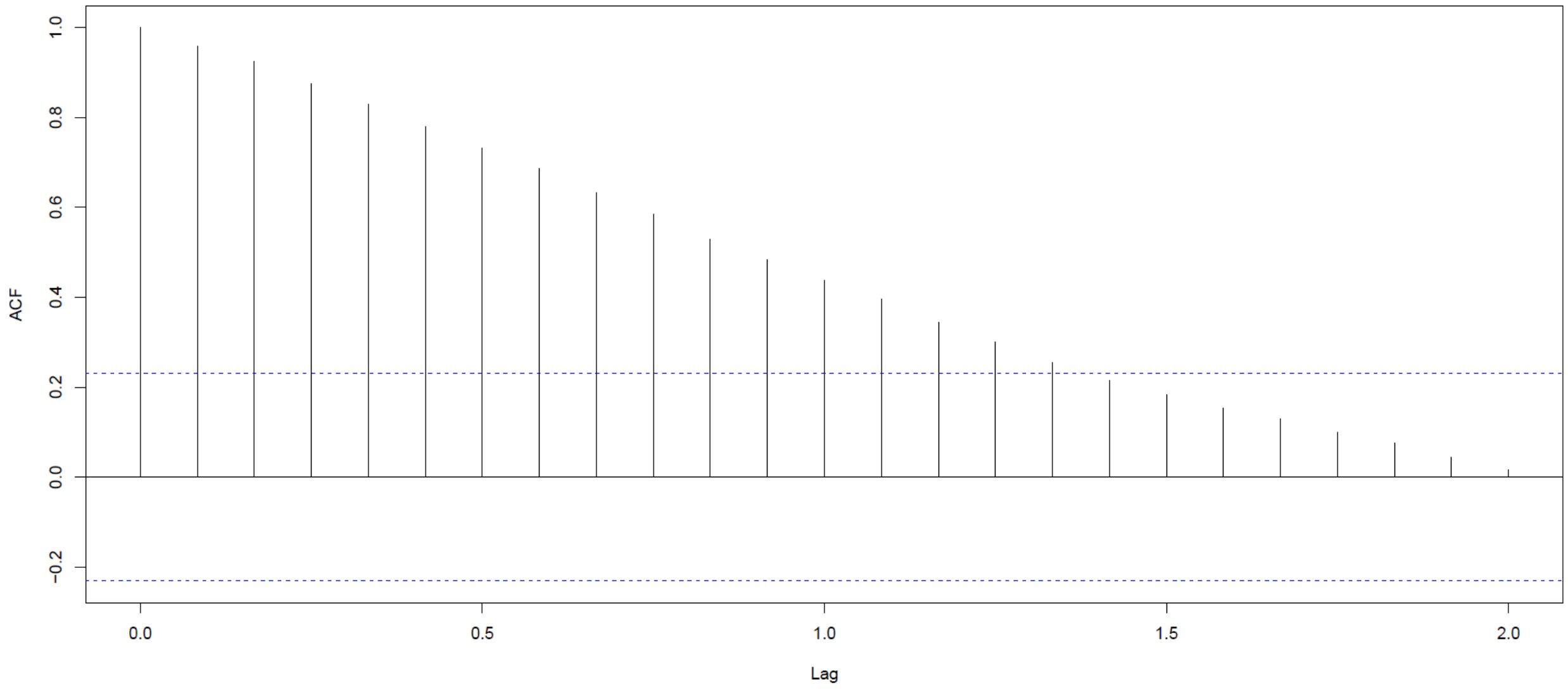

Figure 8. Auto correlation function of the healthcare time series (Period: Jan 2010 – Dec 2015)

**Observation:** It may be observed from Table 6 that the minimum error value is 0.03 per cent which was observed in the month of August 2015, while the maximum error of 9.77 per cent was observed in the month of December 2016. It is also interesting to note that from May 2016 till December 2016, the predicted value remained constant. The RMSE value of this method is 740 that is just 4.71 per cent of the mean value of the actual index, which is 15713. This indicates that this method has been extremely accurate in forecasting the index values of the healthcare time series for the year 2016.

Table 6: Computation results using Method V of forecasting

| Month (A) | Actual Index (B) | Forecasted Index (C) | % Error (C-B)/B *100 | RMSE |
|---|---|---|---|---|
| Jan | 16305 | 15630 | 4.14 | **740** |
| Feb | 15208 | 16252 | 6.86 | |
| Mar | 15149 | 16154 | 6.63 | |
| Apr | 15582 | 16169 | 3.77 | |
| May | 15246 | 16167 | 6.04 | |
| Jun | 15493 | 16167 | 4.35 | |
| Jul | 16299 | 16167 | 0.81 | |
| Aug | 16162 | 16167 | 0.03 | |
| Sep | 16181 | 16167 | 0.09 | |
| Oct | 16472 | 16167 | 1.85 | |
| Nov | 15734 | 16167 | 2.75 | |
| Dec | 14728 | 16167 | 9.77 | |

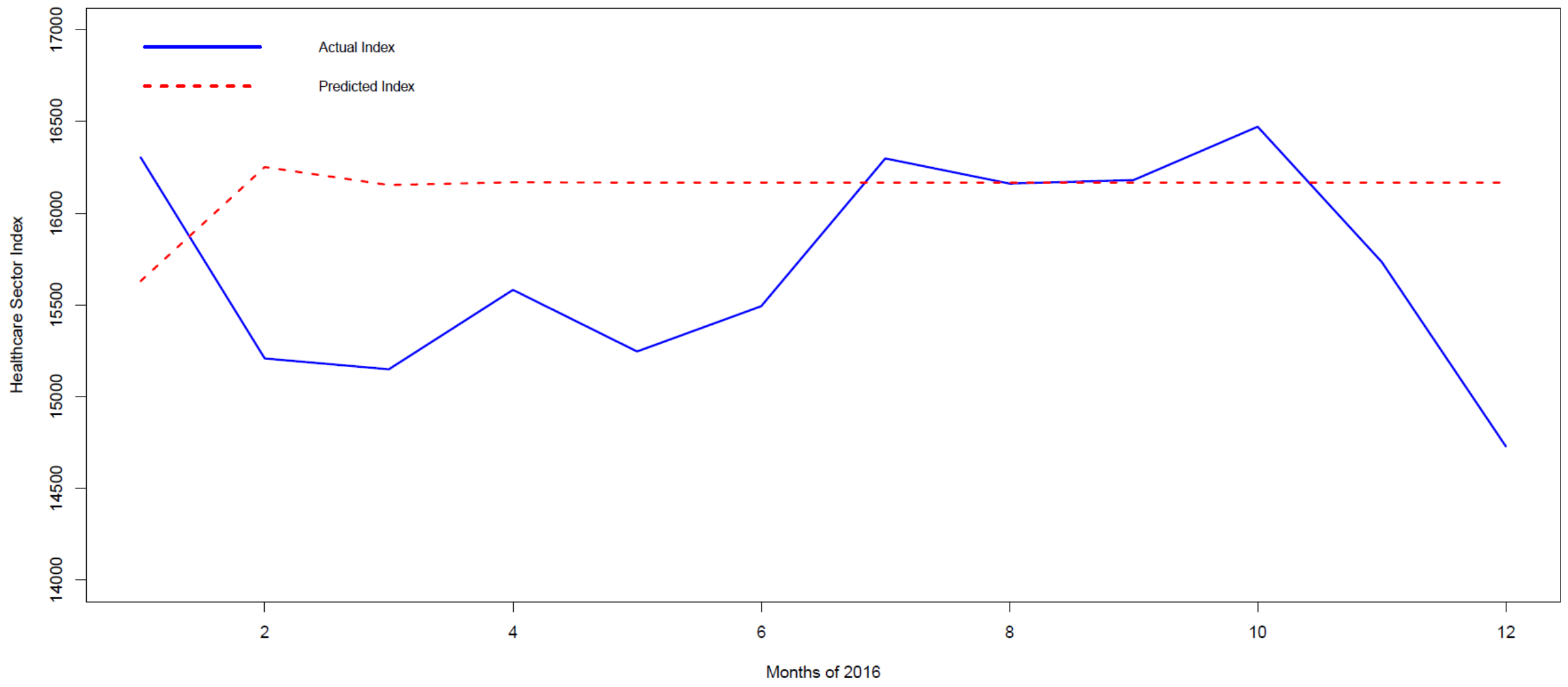

Figure 9. Actual and predicted values of healthcare sector index using Method V of forecasting (Period: Jan 2016 – Dec 2016)

**Method VI:** As discussed earlier in this Section, Method VI is a forecasting method that is based on ARIMA with forecast horizon of 1 month. The ARIMA model used in forecasting is constructed every time before a forecast is made due to change in the training data set for constructing the model. Table 7 presents the forecasting results along with forecast error for each month of the year 2016. A combined RMSE value is also computed for this method as in all other methods. Figure 10 depicts the actual and predicted index values for each month of 2016.

Table 7: Computation results using Method VI of forecasting

| Month (A) | Actual Index (B) | Forecasted Index (C) | % Error (C-B)/B *100 | RMSE |
|---|---|---|---|---|
| Jan | 16305 | 15630 | 4.14 | **800** |
| Feb | 15208 | 16949 | 11.45 | |
| Mar | 15149 | 15014 | 0.89 | |
| Apr | 15582 | 14523 | 6.80 | |
| May | 15246 | 15539 | 1.92 | |
| Jun | 15493 | 15553 | 0.39 | |
| Jul | 16299 | 15405 | 5.48 | |
| Aug | 16162 | 16341 | 1.11 | |
| Sep | 16181 | 16438 | 1.59 | |
| Oct | 16472 | 16136 | 2.04 | |
| Nov | 15734 | 16462 | 4.63 | |
| Dec | 14728 | 15920 | 8.09 | |

**Observations:** From Table 7, it can be observed that the smallest value of the error is 0.39 per cent that was obtained in the month of June 2016, while the highest value of error is 11.45 per cent obtained in the month of February 2016. The RMSE value of the method is 800, which is only 5.09 per cent of the mean of the index values for the year 2016. This indicates that the method is highly accurate in forecasting the monthly index values for the healthcare sector data for the year 2016. Figure 10 also clearly depicts that the difference between the actual and the predicted values of the index has been quite small for all the months of 2016, with only the month of February having an error percentage that is more than 10 per cent. The month of February 2016 has an error per cent of 11.45 in its forecasted index.

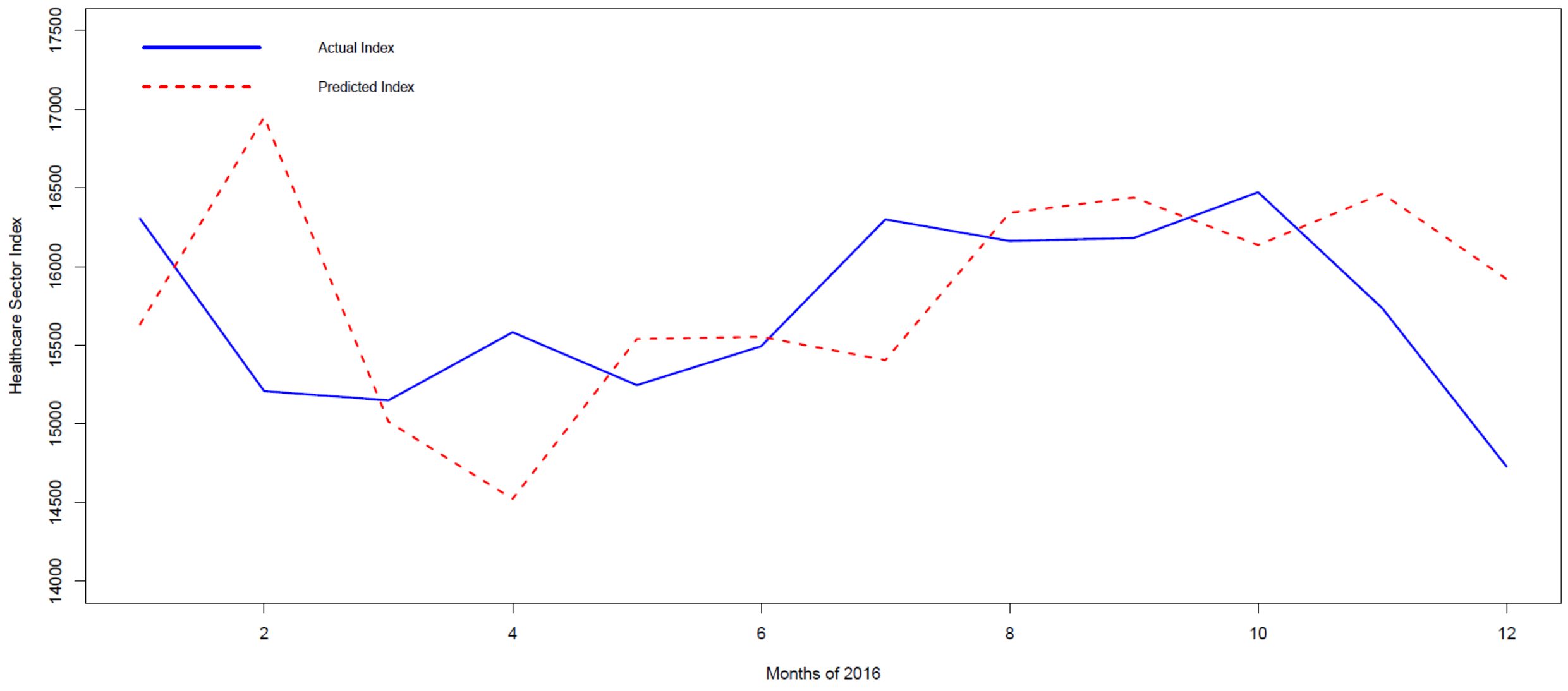

Figure 10. Actual and predicted values of healthcare sector index using Method VI of forecasting (Period: Jan 2016 – Dec 2016)

### Summary of Forecasting Results

In Table 8, we summarize the performance of the six forecasting methods that we have used. For the purpose of comparison between these methods, we have chosen five metrics: (i) minimum (Min) error rate, (ii) maximum (Max) error rate, (iii) mean error rate, (iv) standard deviation (SD) of error rates, and (v) root mean square error (RMSE). Since RMSE is widely considered as a single metric for comparing several approaches in forecasting, we consider that method as the best which has yielded lowest value of RMSE. From Table 8, we observe that Method V that uses ARIMA approach with a forecast horizon of 12 months has performed the best among all the six method with its RMSE value of 740. Method VI that uses ARIMA with forecast horizon of 1 month and Method II using HoltWinters forecasting method with a forecast horizon of 1 month having their RMSE values 800 and 894 respectively are close second and third in ranking in terms of their forecasting accuracies. Method I that uses HoltWinters forecasting with a forecasting horizon of 12 months has yielded highest value of RMSE, and hence its performance has been worst. However, the RMSE value of 2863 is 18.22 per cent of the mean value of the actual index for the year 2016, the mean value of actual index being 15713. This indicates that even though Method I has yielded the highest value of RMSE and hence has performed worst, the forecast accuracy of this method is quite acceptable.

Table 8: Comparison of the performance of the six forecasting methods

| Metrics / Methods | Min Error | Max Error | Mean Error | SD of Errors | RMSE |
|---|---|---|---|---|---|
| **Method 1** | 8.71 | 31.32 | 17.73 | 5.29 | 2863 |
| **Method II** | 0.07 | 10.65 | 4.46 | 3.84 | 894 |
| **Method III** | 0.05 | 19.56 | 8.97 | 7.06 | 1782 |
| **Method IV** | 0.2 | 13.37 | 6.81 | 4.34 | 1339 |
| **Method V** | 0.03 | 9.77 | 3.92 | 3.02 | 740 |
| **Method VI** | 0.39 | 11.45 | 4.04 | 3.41 | 800 |

## 7. Conclusion

In this work, we presented a time series decomposition-based approach for understanding the behavior of the time series of the healthcare sector of the Indian economy during the period January 2010 till December 2016. We have used functions in the R programming language to decompose the time series values into three components- trend, seasonal, and random. The decomposition results have provided us with several valuable insights into the behavior exhibited by the healthcare sector time series during the period under our study. Based on the results, we have been able to identify the months during which the seasonal component in the healthcare time series plays a major role. It has been observed that while the month of October experiences the highest seasonality in the healthcare sector, for the month of February the seasonal effect is lowest. We have also been able to gain an insight into the trend and randomness exhibited by the healthcare time series during the period of our interest. Using these decomposition results, we have proposed six different methods of forecasting the index values of the healthcare sector. It has been observed that while ARIMA-based approach with a forecast horizon of 12 months is the best method of forecast for the time series with its lowest RMSE value, the HoltWinters method with a forecast horizon of 12 months turned to be the worst method exhibiting highest value of RMSE. Since the time series in the year 2016 experienced quite a lot number of ups and downs, the ARIMA method with a forecast horizon of 1 month performed inferior to the ARIMA method with a forecast horizon of 12 months. Even in presence of a random component and a sharply changing trend values in time series values, our techniques have been able to achieve quite an acceptable level of forecasting accuracies.

The results obtained from the above analysis is extremely useful for portfolio construction. When we perform this analysis for other sectors as well, it will help portfolio managers and individual investors to identify which sector, and in turn which stock, to buy/sell in which period. It will also help in identifying which sector, and hence which stock, is dominated by the random component and thus is speculative in nature.